\documentclass[default,sn-mathphys,Numbered]{sn-jnl}

\usepackage{graphicx}%
\usepackage{multirow}%
\usepackage{amsmath,amssymb,amsfonts}%
\usepackage{amsthm}%
\usepackage{mathrsfs}%
\usepackage[title]{appendix}%
\usepackage{xcolor}%
\usepackage{textcomp}%
\usepackage{manyfoot}%
\usepackage{booktabs}%
\usepackage{algorithm}%
\usepackage{algorithmicx}%
\usepackage{algpseudocode}%
\usepackage{listings}%
\usepackage[T1]{fontenc}%

\theoremstyle{thmstyleone}%
\theoremstyle{thmstyletwo}%

\theoremstyle{thmstylethree}%

\begin{document}
\title[Article Title]{Nonlinear improvement of measurement-device-independent quantum key distribution using multimode quantum memory}






\author*[1,2]{\fnm{Yusuke} \sur{Mizutani}}\email{qvuaalnlteuym@gmail.com}

\author[1,2]{\fnm{Tomoyuki} \sur{Horikiri}}\email{horikiri-tomoyuki-bh@ynu.ac.jp}

\affil*[1]{Yokohama National University, 79-5 Tokiwadai, Hodogaya, Yokohama 240-8501, Japan}

\affil[2]{LQUOM, Inc., 79-5 Tokiwadai, Hodogaya, Yokohama 240-8501, Japan}


\abstract{This paper proposes a quantum key distribution (QKD) scheme for measurement-device-independent QKD (MDI-QKD) utilizing quantum memory (QM), which is based on two distinct functions of QM: on-demand storage and multimode storage. We demonstrate a nonlinear increase in the secure key rate due to the utilization of QM. In the protocol incorporating on-demand storage, it is acknowledged that the secure key rate is scaled by $R=O(\sqrt{\eta_{ch}})$ as $\eta_{ch}$, while as an alternative approach, we reveal that the improvement is $O(m_s^2)$, with $m_s$ being the number of modes in frequency (spatial) multiplexing in the scheme incorporating multimode storage. We adopt an atomic frequency comb as a QM that incorporates the two functions and propose an architecture based on MDI-QKD to attain experimental feasibility. This scheme can be extended to quantum repeaters, and even for a single quantum-repeater node, there is a nonlinear enhancement and an experimental incentive to increase the number of modes.}




\maketitle

\section{Introduction}\label{sec1}

Quantum key distribution (QKD) can distribute secure key bit strings between two parties using information-theoretic security. In recent years, transmission distances have experimentally increased, with demonstrations distances exceeding 800 km \cite{wang2022twin}.

However, theoretical security proofs rely on assumptions about the security of various devices, which may not accurately reflect the conditions of actual experiments. Fortunately, there are also implementation papers that bridge the gap between theory and experiment for device imperfections \cite{zhang2022device,mdi-exp-nad,liu2022toward}. Among them, measurement-device-independent quantum key distribution (MDI-QKD) \cite{lo2012measurement}, which can be classified as a meet-in-the-middle (MM) architecture \cite{qr-design}, is particularly noteworthy. This MM-type architecture has yielded numerous leading candidates for both QKDs and quantum repeaters and has achieved theoretical and experimental breakthroughs \cite{zeng2022mode,afc-repeater,herald-dlcz,telcom-dlcz,pompili2021realization,fan2022robust}. 

In the original MDI-QKD scheme, the key rate is strictly limited by the Pirandola--Laurenza--Ottaviani--Banchi (PLOB) bound $R=-\log_2 (1-\eta_{ch})\simeq\eta_{ch}/\ln2\simeq1.44\eta_{ch}$ for $\eta_{ch}\ll1$ \cite{pirandola2017fundamental}, where $\eta_{ch}$ represents the channel transmittance rate between the two users.
To exceed the PLOB limit in QKD, twin-field QKD (TF-QKD) \cite{lucamarini2018overcoming} and its variants, sending or not sending QKD \cite{wang2018twin}, phase-matching QKD \cite{ma2018phase,zeng2020symmetry}, and mode-pairing QKD using asynchronous two-photon interference \cite{xie2022breaking,zeng2022mode} have been proposed. Additionally, memory assisted MDI-QKD (MA-MDI-QKD) \cite{ma-mdi, abruzzo2014measurement} has been proposed and evaluated with quantum memory (QM) in various physical systems \cite{ma-mdi-sps}. These architectures increase the key rate up to $O(\sqrt{\eta_{ch}})$. Moreover, MA-MDI-QKD can be extended as an element of a single quantum-repeater node via improvement of QMs.
Additionally, when analyzing/comparing the performance of quantum repeaters, it is useful to benchmark with the key rate in the QKD protocol \cite{rozpkedek2018parameter,schmidt2020memory}.

Beginning with the previously reported concept \cite{qr-start} of quantum repeaters, which are essential for long-distance quantum communication, various applications can be realized through the entanglement distribution, including QKD, distributed quantum computation \cite{buhrman2003distributed}, and world clocks \cite{komar2014quantum}. Additionally, the proof-of-concept for a quantum internet---a communication network utilizing information-theoretic security---is being developed. It is crucial to develop quantum communication protocols that are ready for connection to the physical layer and higher layers of the network \cite{wehner2018quantum,dahlberg2019link}.

The Duan--Lukin--Cirac--Zoller (DLCZ) protocol realizes a practical experimental implementation of quantum repeaters \cite{duan2001long} and has been implemented in several physical systems, such as quantum dots \cite{stockill2017phase}, trapped ions \cite{stephenson2020high}, and protocols utilizing nitrogen-vacancy centers \cite{pompili2021realization}, as well as entanglement distribution using atomic frequency comb (AFC) QM in recent years \cite{herald-dlcz,telcom-dlcz}.
However, in the realistic protocols realized in the aforementioned experiments, multiplexing is necessary to achieve practical rates (on the order of Hz or higher) \cite{collins2007multiplexed,qr-guha,qr-krovi}.
Therefore, our focus has been on the multimode storage of QM, and we propose a scheme that introduces new possibilities.

We adopt AFC as a QM \cite{afzelius2009multimode} comprising two functions: multimode storage and on-demand storage. These two functions have a distinct impact on the key rate.
To this end, we propose an experimentally implementable protocol, which is a one-element MDI type protocol, to analyze the impact of multimode QM on the key rate.

There are two primary methods for achieving a key rate of $O(\sqrt{\eta_{ch}})$. The first is to utilize a one-photon interference scheme, as exemplified by proposals such as TF-QKD \cite{lucamarini2018overcoming} and mode-pairing schemes that alleviate phase-locking issues in one-photon interference \cite{xie2022breaking,zeng2022mode}. The second is a scheme akin to MA-MDI-QKD, in which QM allows Bell state measurements (BSMs) to be performed even if photons arrive asynchronously at the central interference system. As a practical proposal, protocols utilizing on-demand storage can achieve $O(\sqrt{\eta_{ch}})$ within a certain range of distances, contingent upon the decoherence time $T_2$ of the QM \cite{ma-mdi,luong2016overcoming}.

Herein, we present an alternative approach, whereby we demonstrate that the key rate can be increased by $O(m_s^2)$, with $m_s$ representing the number of modes, through the incorporation of frequency multiplexing and frequency shifting in the middle BSM. Even after normalization by the number of modes, it is possible to benefit from frequency multiplexing. 

The remainder of this paper is organized as follows. Section \ref{sec:system} describes the structure of the protocol incorporating the functions of QM and proposes a feasible setup for the mechanism. In Section \ref{sec:keyrate}, in addition to the on-demand storage of QM, the multimode storage of the frequency and spatial modes is incorporated into the calculation and analyzed as the key rate of the one-element MDI-QKD protocol. Finally, Section \ref{sec:conc} summarizes the study and discusses future extensions to quantum-repeater networks. 

\section{System description}\label{sec:system}

In this section, we first provide an overview of the fundamental principles of the on-demand/multimode protocol, which is predicated on the functionality of QM. Subsequently, we comprehensively explain the protocol.

\subsection{Protocol description}\label{section:protocol-detail}

\begin{figure}
\centering
\includegraphics[width=100mm]{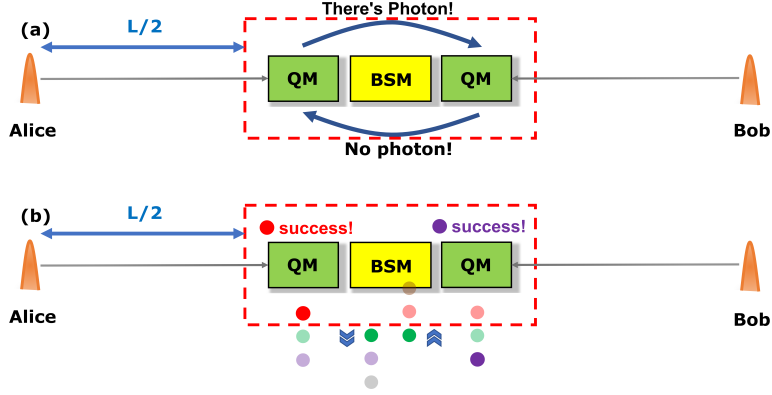}
\caption{Schematic of the system of protocols incorporating QM functions, specifically on-demand and multimode storage. (a) On-demand protocol: We assume the presence of a QM with on-demand storage. This function allows the QM on Alice's (Bob's) side to determine whether a photon has been successfully absorbed and obtain its stored information. (b) Multimode protocol: A QM with frequency multiplexing is employed. The frequency mode of the photon that is successfully absorbed on Alice’s (Bob’s) side is read from the QM and shifted to a fixed frequency mode (depicted as green in the figure). The frequency mode used in the central BSM is consistently fixed.}

\label{fig:on-mm-protocol}
\end{figure}

We present a system of protocols that incorporates the functions of QM, specifically on-demand and multimode storage. The first protocol incorporates a QM with on-demand storage, as depicted in Fig.\ref{fig:on-mm-protocol}(a), and the second protocol incorporates a QM with frequency (spatial) multimode storage, as depicted in Fig.\ref{fig:on-mm-protocol}(b).\\

\noindent {\bf On-demand QM protocol.}\\

First, we describe a protocol that incorporates a QM with on-demand storage, as illustrated in Fig.\ref{fig:on-mm-protocol}(a). The QM on Bob’s (Alice’s) side is notified of whether a photon has been successfully absorbed by the QM on Alice’s (Bob’s) side, allowing the determination of whether the photonic quantum state has been stored in the QM on Alice’s (Bob’s) side. If the presence of the photon is heralded by the successful loading of both QMs, the middle BSM is executed as usual. However, if the loading of the QM on Alice’s (Bob’s) side is unsuccessful, the QM on Bob’s (Alice’s) side stores the photon information and awaits the next successful loading of the QM on Alice’s (Bob’s) side.\\

\noindent {\bf Multimode QM protocol.}\\

We next describe the protocol incorporating a QM with frequency-multiplexing storage, as illustrated in Fig.\ref{fig:on-mm-protocol}(b). The protocol utilizes a QM with the storage of frequency multiplexing, as demonstrated in the AFC \cite{afzelius2009multimode}. The frequency mode of the photon successfully absorbed on Alice’s (Bob’s) side is retrieved from the QM and shifted to a fixed frequency mode with efficiency $\eta_{f}$. The frequency mode utilized in the middle BSM is consistently fixed \cite{saglamyurek2014integrated}. This protocol does not require on-demand storage and can be implemented with fixed-time storage.\\

\subsection{Protocol setup}

Here, we detail the implementation of the scheme depicted in Fig.\ref{fig:on-mm-protocol} utilizing QM.

\begin{figure}[h]
\centering
\includegraphics[width=100mm]{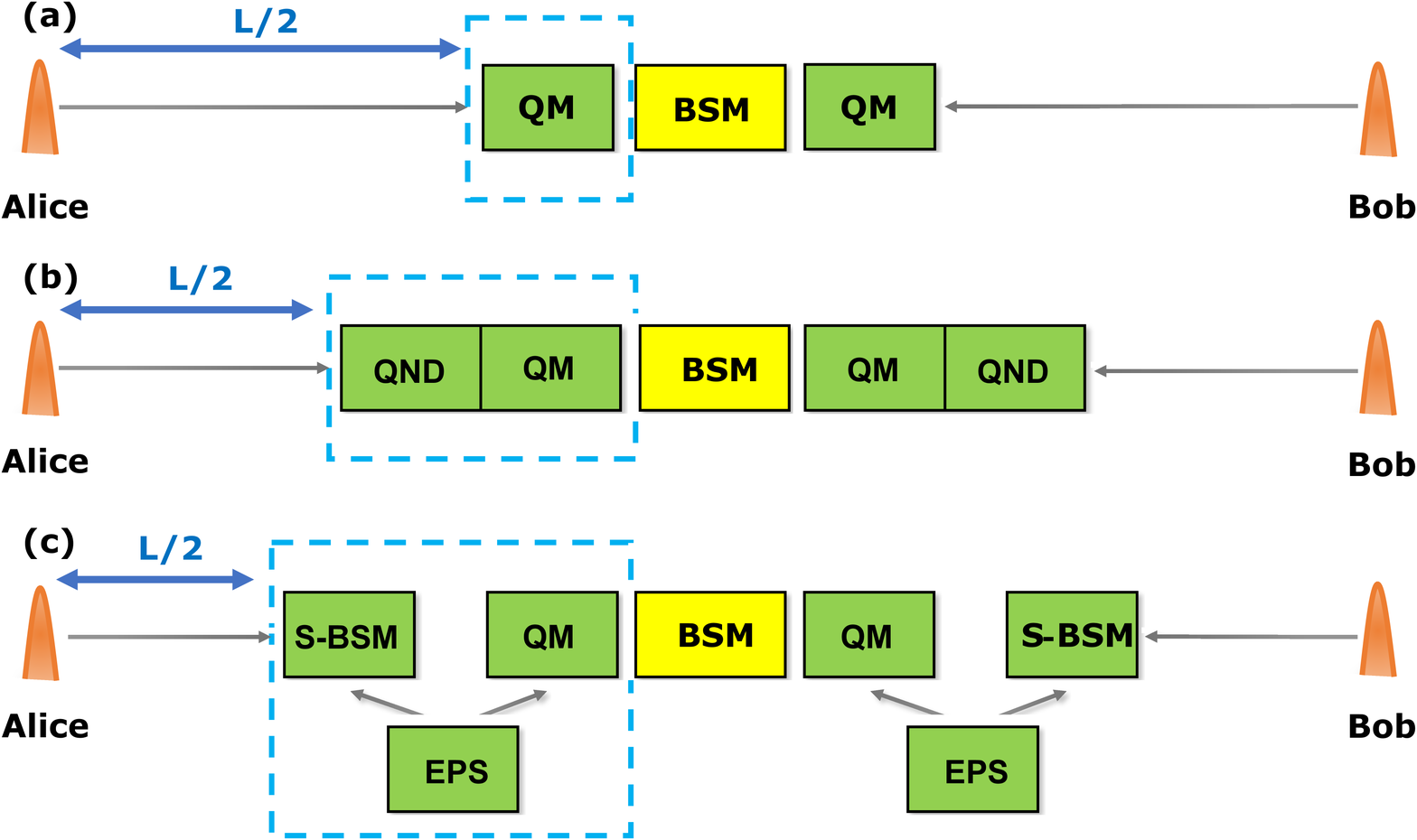}
\caption{In (a), the original MA-MDI-QKD protocol proposed in \cite{ma-mdi} is presented. In (b), a heralding scheme utilizing quantum non-demolition (QND) measurement is presented. The QND nondestructively heralds the presence of a photon, indicating successful absorption into the QM. Once both QMs have completed their readings, the stored state of the QMs is retrieved as photons, and a BSM is executed in the middle. In (c), a heralding scheme utilizing BSM is presented. In each round, entangled photons from the entangled photon source (EPS) are sent to the QM and a side-BSM (S-BSM), where the QM stores the photon information and the S-BSM performs a BSM with photons from Alice (Bob). If it is successful, the heralding of absorption of the photon stored in the QM is successful, the quantum state is taken as a photon from both QMs, and a BSM is performed in the middle.}
\label{fig:protocol}
\end{figure}

\begin{figure}
\centering
\includegraphics[width=100mm]{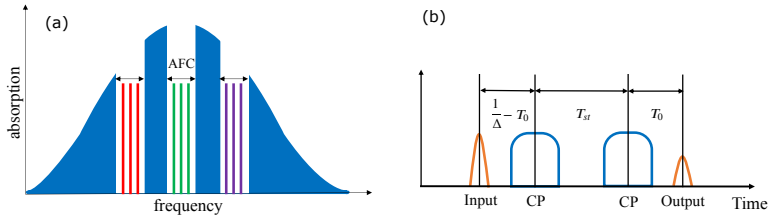}
\caption{Process of writing and reading on-demand and multimode AFC QM is illustrated in Fig.\ref{fig:on-mm-protocol}. (a) Structure of an AFC with frequency-multiplexing storage, where photons are absorbed. (b) Write and read processes are depicted on the time axis. The interval between the peaks of the comb, which is referred to as $\Delta$, depends on the structure of the AFC. A pulse is absorbed by the AFC as an input mode, followed by the application of a control pulse (CP). After a period of $T_{st}$, a second CP is applied, and photons are regenerated as the output mode after a total of $1/\Delta+T_{st}$.}
\label{fig:afc}
\end{figure}

In Fig.\ref{fig:on-mm-protocol}, the three functions required of the QM are the function of heralding, which nondestructively transmits the arrival of a photon; the on-demand storage; and the multimode storage. Specifically, in Fig.\ref{fig:on-mm-protocol}(a), the QM requires the on-demand storage and the function of heralding, and in Fig.\ref{fig:on-mm-protocol}(b), the QM requires the multimode storage and the function of heralding. However, there is currently no practical QM that possesses all three of these functions. Nevertheless, AFC QM has two of these functions: on-demand and multimode storage. Therefore, when utilizing AFC QM, which is suitable for multimode storage, it is necessary to incorporate the function of heralding separately in actual experiments. Fig.\ref{fig:protocol} illustrates a scheme with heralding. Fig.\ref{fig:protocol}(a) depicts the basic MA-MDI-QKD scheme \cite{ma-mdi}, and we extend the QM in this protocol by replacing it with an AFC.

Fig.\ref{fig:afc} illustrates the physical system that allows the on-demand and multimode storage of the AFC QM. As depicted in Fig.\ref{fig:afc}(a), multiple AFCs can be created through inhomogeneous broadening, allowing absorption of photons of different modes. Additionally, as illustrated in Fig.\ref{fig:afc}(b), the storage time $T_{st}$ can be set to an arbitrary value through the timing of CP irradiation. However, the AFC alone cannot herald the arrival of a photon sent by Alice (Bob) without destroying the quantum information.

There are three conventional methods for heralding the arrival of a photon. The first is heralding with the QM alone, as shown in Fig.\ref{fig:protocol}(a). This method has been demonstrated using spin--photon interfaces, e.g., silicon--vacancy centers \cite{bhaskar2020experimental}. The second method is heralding with QND, as depicted in Fig.\ref{fig:protocol}(b). This approach incorporates QND \cite{afc-qnd} to realize the function of heralding external to the QM when the QM does not possess this function. However, QND in AFC has only been achieved at the classical light level---not at the single-photon level. As an alternative, heralding can be realized indirectly by incorporating S-BSM and EPS in Fig.\ref{fig:protocol}(c). This method allows the determination of whether a photon has been absorbed into the QM based on the outcome of the BSM, making it useful for ensemble-based QMs with multimode storage, as it is difficult to achieve the function of heralding with such QMs.\\

\subsubsection{Device description}

We now provide an overview of each component, assuming the absence of malicious third-party actors.\\

\textit{Photon source}

In the proposed scheme, Alice and Bob's single-photon source performs the calculation with a weak coherent pulse (WCP) and an optimized average photon number $\mu,\nu$. The utilization of EPSs has been actively studied in recent years. A high-fidelity degenerate EPS with a telecommunication wavelength was reported in a cavity two-photon configuration \cite{niizeki2020two}, as well as a non-degenerate EPS, where the signal and idler are the QM wavelength and telecommunication wavelength, respectively \cite{telcom-dlcz}. In practical examples of protocols using EPSs, linear improvements in the key rate have been achieved through multiplexing with existing EPS devices \cite{mm-eps-merit}. While the EPSs considered in this study do not take into account the probability of two-photon generation in each mode, simulation results indicate that for $p_{2}>0$, a constant $p_{2}$ range in a frequency-multiplexed quantum-repeater scheme has little effect on the key rate \cite{qr-guha}. Alternative options have been proposed to replace quantum dot-based photon sources with low $p_{2}$ values, such as the approach presented in reference \cite{muller2014demand} and the setup outlined in reference \cite{ma-mdi-sps}, which eliminates the EPS.\\

\textit{Quantum memory}

Our scheme requires a QM with on-demand storage and multimode storage. 
With regard to the function of on-demand storage, if photon absorption in Bob's (Alice's) QM is unsuccessful, Alice (Bob) must store the quantum state in the QM until Bob (Alice) achieves a successful absorption. However, during this period, the QM becomes susceptible to dephasing due to $T_2$; the initial state of the QM is $\rho$, and the state of the QM after dephasing for a period of time $t$ can be modeled using the Completely Positive and Trace Preserving map, as reported in \cite{razavi2009quantum}.

\begin{align}
    	\varepsilon(\rho) &= [1-\lambda_{dp}(t)]\rho+\lambda_{dp}(t)Z\rho Z. 
\end{align}

Here,

\begin{align}
    	\lambda_{dp}(t) &= \frac{1-e^{-\frac{t}{T_2}}}{2}, 
\end{align}

where $Z$ represents the Pauli $Z$ operator.
AFC QM, which is based on rare-earth doped crystals, is the first viable candidate for QM with multimode storage. In recent times, AFC QM has progressed to the experimental stage of multimode storage, and in the case of temporal multiplexing, 51-$\mu s$ has been achieved with $\mathrm{^{151}Eu^{3+}{\colon}Y_2SiO_5}$ at 100 modes in fixed-time storage, and 0.541-$ms$ at 50 modes in on-demand storage has been realized. In frequency multiplexing, $\mathrm{Pr^{3+}{\colon}Y_2SiO_5}$ can achieve 100 modes in fixed-time storage \cite{jobez2016towards,ortu2022multimode}. Furthermore, as an experimental example, entanglement distribution using the time-multimode storage has been reported \cite{telcom-dlcz}.\\

\textit{Frequency shifter}

In the scenario where photons of varying frequency modes are to be regenerated, frequency shifting can be accomplished by using an electro-optic modulator to perform a BSM. The technology required for this process is serrodyne frequency shifting, which was demonstrated to have a conversion efficiency of 80\% for a frequency shift of approximately 1 GHz in experiments utilizing an AFC \cite{saglamyurek2014integrated}. \\

\textit{Bell state measurement}

In this study, there is a middle BSM and an S-BSM employed for heralding when a photon is successfully absorbed by the QM. The success probability for the middle BSM can be calculated using $Y_{11}$, as described in Appendix \ref{ap:subsec1}:

\begin{align}
    P^{MBSM} &= Y_{11}(\eta_A,\eta_B), 
\end{align}

where $\eta_A$ and $\eta_B$ represent the probabilities of photons reaching the middle BSM on Alice’s and Bob’s sides, respectively. Similarly, the S-BSM used for heralding, which is depicted in Fig.\ref{fig:protocol}(c), can be viewed as an asymmetric MDI-QKD system:

\begin{align}
    P^{SBSM}_K &= Y_{11}(\eta_K,\eta_{ent}\eta_{d}) \quad(K=A,B).
\end{align}

With regard to multiplexing, a BSM utilizing spatial modes \cite{mm-hom} and a BSM utilizing temporal modes \cite{valivarthi2014efficient} have been proposed. Additionally, a BSM setup that exceeds 50\% \cite{grice2011arbitrarily} has been proposed.\\

\section{Key rate analysis}\label{sec:keyrate}

In this section, we calculate the key rate for the proposed scheme illustrated in Fig.\ref{fig:protocol}(c). In QKD experiments, it is commonly assumed that normal operation proceeds without interference from eavesdroppers and is only affected by system imperfections.
Under these conditions, in the setting of an infinite key, the key rate in Fig.\ref{fig:protocol} setting is bound by the following:

\begin{align}
    R &= Y_{11}^{QM}[1-h(e_{X})-fh(e_{Z})], 
\end{align}

where $e_{X}$ and $e_{Z}$ represent the qubit error rates (QBERs) between Alice and Bob in the X and Z bases, respectively, and $Y_{11}^{QM}$ represents the rate at which raw key bits are generated. Additionally, $h(p)=-p\log_2(p)-(1-p)\log_2(1-p)$ is the Shannon's binary entropy function.
In this study, inefficiency is analyzed with $f=0$, and other parameters are presented in Table \ref{tab:table1}.\\

\begin{table}[b]
\caption{\label{tab:table1}%
Parameters associated with the QKD protocol discussed in Section \ref{sec:keyrate}.
}
\begin{tabular}{@{}llll@{}}
\toprule
\textrm{Parameter}&
\textrm{Symbol}\\
\midrule
rate at which the QMs and BSM are successful & $Y_{11}^{QM}$\\
probability of a successful BSM & $Y_{11}$\\
QBER in basis $S={X,Z}$ & $e_S$\\
QM reading efficiency & $\eta_r$\\
frequency-shifting efficiency & $\eta_f$\\
detector efficiency & $\eta_d$\\
channel-loss efficiency & $\eta_{ch}$\\
entangling efficiency & $\eta_{ent}$\\
probability of successfully loading a QM $K = {A,B}$ & $\eta^{mm}_{K}$\\
optical coherence time & $T_2$\\
number of modes in spectral/spatial multimode & $m_s$\\
number of modes in temporal multimode & $m_t$\\
attenuation length & $L_{att}$\\
dark count rate per pulse & $p_{dc}$\\
\botrule
\end{tabular}
\end{table}

\noindent {\bf On-demand AFC protocol.}\\

\begin{figure}[h]
\centering
\includegraphics[width=95mm]{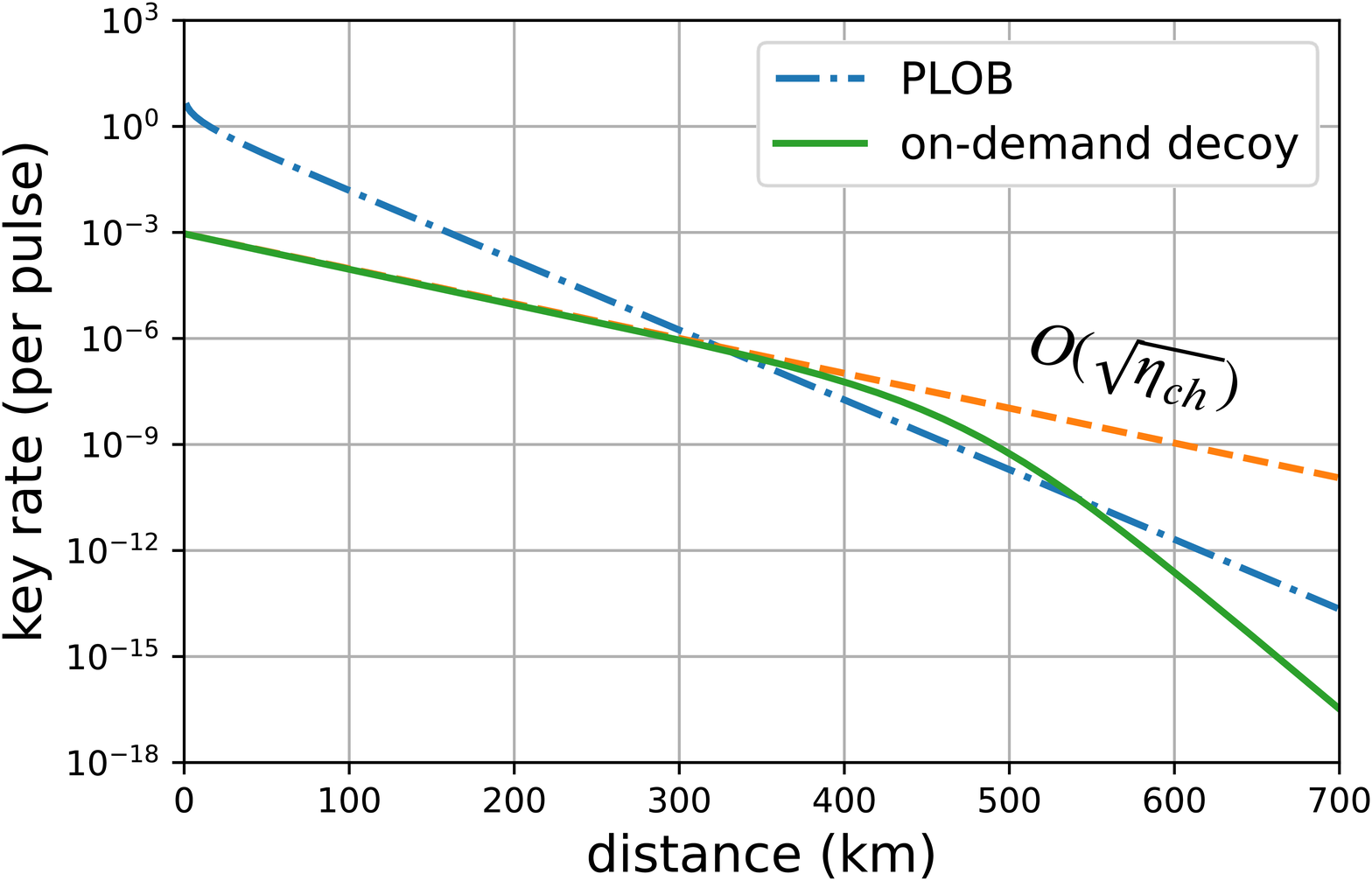}
\caption{Comparison of the key rate of the on-demand AFC protocol with the PLOB bound (R=1.44$\eta_{ch}$). The parameters were set as follows: $L_{att}=22, \eta_{d}=0.93, \eta_{ent}=0.5, p_{dc}=10^{-8}, T_2=300\times10^{-6}$.}
\label{fig:afc-ondemand-protocol}
\end{figure}

As the protocol depicted in Fig.\ref{fig:on-mm-protocol}(a) necessitates the utilization of QM with the function of heralding, we realized an on-demand protocol utilizing AFC QM by integrating S-BSM, as shown in Fig.\ref{fig:protocol}(c).
The MDI-QKD scheme with on-demand QM has been analyzed in previous studies. In \cite{abruzzo2014measurement}, decoherence due to time evolution of the QM was not considered, and in \cite{ma-mdi}, decoherence was considered.
As depicted in Fig.\ref{fig:afc-ondemand-protocol}, the key rate of the protocol scales with $O(\sqrt{\eta_{ch}})$.  This scaling appears to be valid over a limited distance range up to 400 km.
The reason is that the effects of QM decoherence become increasingly dominant as the distance between the QM and Alice or Bob increases, ultimately causing the key rate to fall below the PLOB limit.
The average storage time required for the QM can be calculated using the equation $T_{st} = E\{N_A-N_B\}\tau=\frac{2(1-\eta)\tau}{\eta(2-\eta)} \approx \frac{\tau}{\eta} \quad(\eta \ll 1)$ \cite{ma-mdi}, where $N_A$ and $N_B$ are the probability variables for successful loading of the QM on Alice’s and Bob’s sides, respectively. 
Here, $\tau$ represents the repetition period. $\eta$ represents the loading efficiency of the QM, with $\eta_A$ and $\eta_B$ corresponding to Alice’s and Bob's sides, respectively, and $\eta=\eta_A=\eta_B$. Instead of calculation based on the average storage time, an alternative approach is to optimize the number of modes in temporal multiplexing $m_t$ according to distance, resulting in the scaling of $R=O(\sqrt{\eta_{ch}})$, as detailed in Appendix \ref{ap:subsec2}.\\

\noindent {\bf Multimode AFC protocol}\\

\begin{figure}[h]
\centering
\includegraphics[width=100mm]{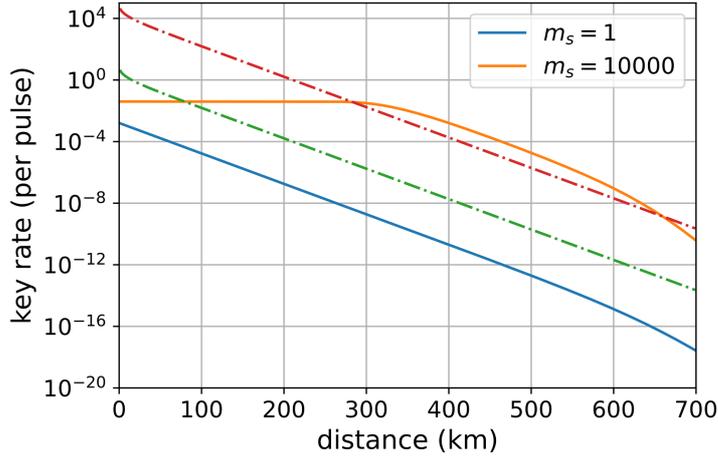}
\caption{Key rate of the multimode AFC protocol is compared with the PLOB bound (R = 1.44$m_s\eta_{ch}$). Simulations were performed with $m_s=1,10000$. The parameters used in the simulations were $L_{att}=22$ km, $\eta_{d}=0.93$, $\eta_{ent}=0.5$, $p_{dc}=10^{-8}$, and $\eta_{f}=0.8$. The PLOB bound is depicted as green (lower dashdot) and red (upper dashdot) for $m_s = 1$ and $1000$, respectively.}
\label{fig:afc-mm-protocol}
\end{figure}%

\begin{figure}[h]
\centering
\includegraphics[width=100mm]{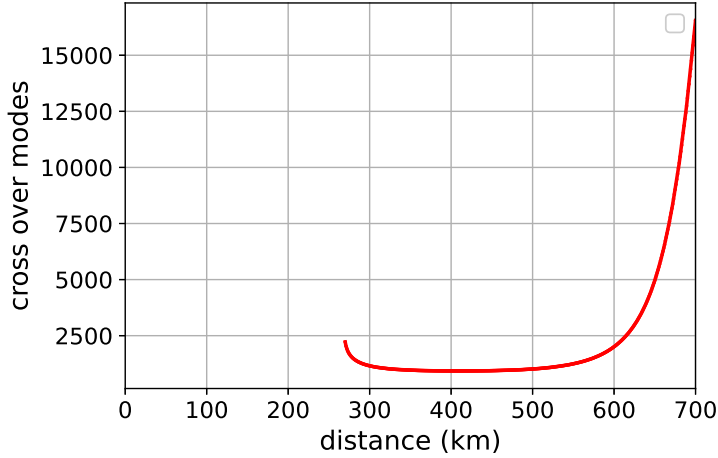}
\caption{Number of modes at which the key rate of the multimode protocol exceeds the PLOB limit at each distance (crossover modes).
Until the distance reaches approximately 270 km, it is infeasible to surpass the PLOB limit. However, beyond this range, it becomes possible to exceed the PLOB limit by utilizing approximately 1000 modes, within the practical distance range.}
\label{fig:cross-over}
\end{figure}%

We realized the multimode protocol utilizing the multimode storage of the AFC QM by integrating S-BSM, as shown in Fig.\ref{fig:protocol}(c). When QM is utilized with multimode storage via frequency or spatial multiplexing, $Y_{11}^{QM}$ is expressed as follows:

\begin{align}\label{eq:7}
    Y_{11}^{QM} &= \mu\nu e^{-\mu-\nu}Y_{11}(\eta_A^{mm},\eta_B^{mm}), 
\end{align}

where $\mu, \nu$ represent the average photon numbers of the WCPs on Alice’s and Bob’s sides, respectively; $\eta_K^{mm} (K=A,B)$ represents the success probability of S-BSM in multiplexing, which is given below; and $m_s$ represents the number of modes in frequency or spatial multiplexing.

\begin{align}
    \eta_K^{mm} &= 1-(1-Y_{11}(\eta_{ch}(L_K)\eta_d,\eta_{ent}\eta_d))^{m_s} \quad(K=A,B)
\end{align}

\begin{figure}
\centering
\includegraphics[width=100mm]{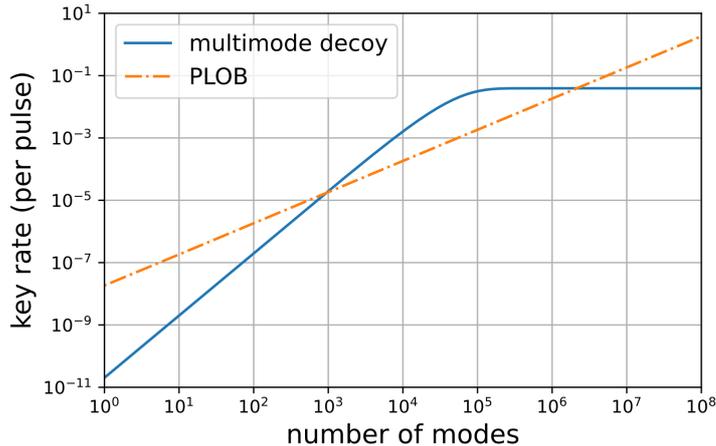}
\caption{At a distance of 400 km, the key rate of the multimode AFC protocol depends on the number of modes. Specifically, the rate increases by $O(m^2_s)$ as the number of modes $m_s$ increases. Furthermore, when 1000 modes or more are utilized, the multimode AFC protocol surpasses the PLOB bound.}
\label{fig:keyrate-modes}
\end{figure}%

The incorporation of multiplexing in the S-BSM is considered, where $\eta_d, \eta_{ent}$ represent the detector efficiency and entanglement source efficiency, respectively. The mathematical framework for the analysis of spatial multiplexing is equivalent to that of frequency multiplexing. Despite the experimental differences between frequency and spatial multiplexing, 36 modes of transmission have been achieved using a single fiber \cite{mizuno201412}, with ongoing efforts to develop spatial multiplexing in QMs \cite{gundougan2012quantum}.
As shown in Fig.\ref{fig:afc-mm-protocol}, the plateau region became larger with an increase in the number of modes $m_s$. The key rate of the multimode AFC protocol, i.e., the original MDI-QKD protocol, for $m_s=1$ was significantly below the PLOB bound; however, for $m_s=10000$, the key rate of the multimode AFC protocol surpassed the PLOB bound.
As shown in Fig.\ref{fig:afc-mm-protocol}, the key rate increased by $O(m_s^2)$ at long distances as the number of modes in frequency (spatial) multiplexing $m_s$ increased. 
This analysis indicates that the improvement persists even after normalization by the number of modes, providing motivation for experimental efforts to increase the number of modes in frequency (spatial) multiplexing, even in the absence of a quantum repeater with multiple links ($N\geq2$).

In the scheme depicted in Fig.\ref{fig:on-mm-protocol}(b), it should be noted that the QM necessitates only fixed-time storage, rather than on-demand storage.

Fig.\ref{fig:cross-over} illustrates that the number of crossover modes in the multimode protocol is approximately 1000 within the range of 300--500 km.
As depicted in Fig.\ref{fig:keyrate-modes}, the key rate is also equivalent to the PLOB bound before the number of modes reaches 1000, but in $\mathrm{Pr^{3+}\colon Y_2SiO_5}$, the number of modes that can be stored by frequency multiplexing is currently limited to approximately 100. To exceed this number, the utilization of either spatial or temporal multiplexing is necessary.
Theoretically, 100-mode multiplexing is possible through the incorporation of spatial multiplexing \cite{ortu2022multimode}, increasing the total number of modes to $10^4$. 
In this context, the middle BSM is executed in a single mode. Consequently, despite the increase in $\eta_{K}^{mm}$ due to multiplexing, the maximum attainable number of bits in a single round remains limited to 1. Thus, with an increase in the parameter $m_s$, the key rate eventually reaches a plateau, as depicted in Figs.\ref{fig:afc-mm-protocol} and \ref{fig:keyrate-modes}.

\section{Conclusion}\label{sec:conc}

We examined the MA-MDI-QKD configuration utilizing a multimode QM. We analyzed the impact on the key rate. The results indicated that the protocol utilizing the QM with on-demand storage yields a key rate of $O(\sqrt{\eta_{ch}})$, which depends on $T_2$. In comparison, the protocol utilizing multimode storage achieves a key-rate increase of $O(m_s^2)$ through frequency (spatial) multiplexing. Furthermore, with current technology, frequency multiplexing alone yields $m_s\sim100$ in the AFC, but when frequency multiplexing and spatial multiplexing are combined, it is possible to exceed the PLOB bound. This outcome indicates that multiplexing has considerable potential for single-element QKD protocols and is effective for single quantum-repeater node protocols. In the future, advancements in QMs can extend these configurations to scalable quantum repeaters and contribute to rate enhancement through multimode quantum repeaters or QKDs. Additionally, the theoretical and experimental work on these protocols brings us closer to the ultimate goal of the quantum internet, as it serves as a stepping stone between point-to-point QKD and full quantum-repeater networks.\\

\noindent {\bf Acknowledgments}
We thank Mohsen Razavi, Nicolò Lo Piparo, Masoome Fazelian, Kae Nemoto, and Toshihiko Sasaki for valuable discussions. This research was supported by JST Moonshot R\&D Grant Number JPMJMS226C, JSPS KAKENHI Grant Number JP20H02652, and the National Institute of Information and Communications Technology Young Researchers Lab. 
We also acknowledge the members of the Quantum Internet Task Force, which is a research consortium aiming to realize the quantum internet, for comprehensive and interdisciplinary discussions of the quantum internet. \\


\begin{appendices}

\section {Calculation of key rates dependent on \texorpdfstring{$T_2$}{Lg}}\label{ap:subsec1}

If QM with on-demand storage is utilized, $Y_{11}^{QM}$ can be described as follows \cite{ma-mdi,luong2016overcoming}:

\begin{align}
    Y_{11}^{QM} &= \frac{Y_{11}(\eta_{early},\eta_{late})}{N_L(\eta_{\mu A},\eta_{\nu B})}\cdot\frac{\eta_{1A}\eta_{1B}}{\eta_{\mu A}\eta_{\nu B}}\mu\nu e^{-\mu-\nu}, 
\end{align}

where $\eta_{1K}, \eta_{\mu(\nu) K}\quad(K=A,B)$ represent the readout efficiency from QM in single-photon states and average photon number $\mu(\nu)$, respectively. Moreover, $Y_{11}$ is given as follows \cite{ma2012alternative}:

\begin{flalign}
    & Y_{11}(\eta_A,\eta_B) \notag &\\
    &= (1-p_{dc})^2 \times \left[\frac{\eta_{A}\eta_{B}}{2}+(2\eta_{A}+2\eta_{B}-3\eta_{A}\eta_{B})p_{dc}+4(1-\eta_{A})(1-\eta_{B})p_{dc}^2\right], 
\end{flalign}

where $N_L$ represents the average number of attempts to load both QMs.
The average number of attempts to load both memories, i.e., $N_L(\eta_A,\eta_B)$, is approximated by $N_L = \frac{3-2\eta_A}{\eta_A(2-\eta_A)} \approx \frac{3}{2\eta_A}$ when $\eta_A = \eta_B \ll 1$ in the no-multimode scenario \cite{collins2007multiplexed}.
Furthermore, $\eta_{early}$ and $\eta_{late}$ represent the loading efficiencies of the QMs for photons arriving early and late, respectively \cite{ma-mdi}:

\begin{equation}
\eta_K^{QM}= \left \{
\begin{array}{l}
\eta_{late} = \eta_{r0}\eta_d,\quad\text{if QM K is late}, \\
\eta_{early} = \eta_r(t)\eta_d,\quad\text{if QM K is early},
\end{array}
\right.
\end{equation}

where $t$ represents the storage time of the QM in the event that it is loaded early, which is given as $(t=|N_A-N_B|\tau)$. The calculation of the expected value for $\eta_{early}$ is as follows:

\begin{flalign}
    & \overline{\eta_{early}} \notag &\\
    &= \eta_d\eta_{r0}E\{\exp(-|N_A-N_B|\tau/T_2)\} \notag &\\
    &= \frac{\eta_d\eta_{r0}\eta_{A}\eta_{B}}{\eta_{A}+\eta_{B}-\eta_{A}\eta_{B}} \times \left[\frac{1}{1-e^{-\frac{\tau}{T_2}}(1-\eta_A)}+\frac{1}{1-e^{-\frac{\tau}{T_2}}(1-\eta_B)}-1\right].
\end{flalign}

\section {Calculation of key rates dependent on temporal multiplexing \texorpdfstring{$m_t$}{Lg}}\label{ap:subsec2}

\begin{figure}[h]
\centering
\includegraphics[width=90mm]{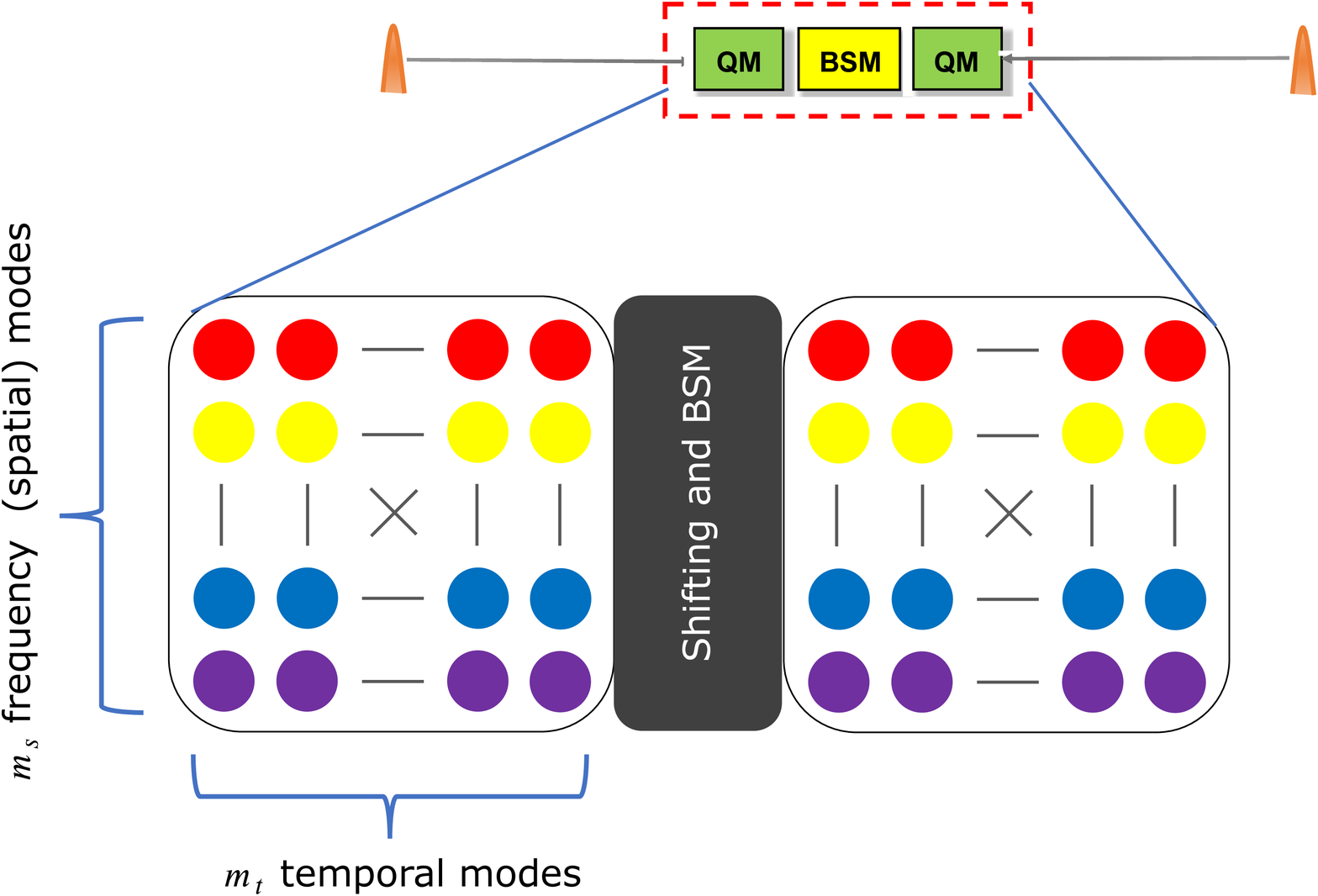}
\caption{Schematic of a QM used in protocols that incorporate both frequency (spatial) multiplexing and time multiplexing. Here, each QM stores $m_s m_t$ photons and performs a frequency/time bin shift in the center for BSMs.}
\label{fig:afc-ms-mt-qm}
\end{figure}

The multimode AFC protocol (frequency multiplexing/spatial multiplexing scheme) in Section \ref{sec:keyrate} was considered as temporal multiplexing $m_t$ with $m_t=1$, but it can be extended to the case of temporal multiplexing ($m_t>1$). A schematic of the QM when temporal multiplexing is incorporated is depicted in Fig.\ref{fig:afc-ms-mt-qm}. It is necessary to store $m_sm_t$ photons across $m_t$ slots and match them accordingly.

\begin{align}
    Y_{11}^{QM} &= \frac{\mu\nu e^{-\mu-\nu}Y_{11}(\eta_A^{mm},\eta_B^{mm})}{m_t}
\end{align}

Here, $Y_{11}$ represents the success rate of the BSM, as described in Appendix \ref{ap:subsec1}. The success probability $\eta_K^{mm}$ of the frequency-multiplexed and time-multiplexed S-BSM is given as follows:

\begin{align}
    & \eta_K^{mm} = 1-(1-Y_{11}(\eta_{ch}(L_K)\eta_d,\eta_{ent}\eta_d))^{m_sm_t} \quad(K=A,B).
\end{align}

Here, in contrast to frequency/spatial multiplexing, temporal multiplexing results in a linear increase in $O(m_t)$, as $m_t$ appears in the denominator of the rate equation \ref{eq:7}, reducing the number of middle BSM attempts per unit time.

\begin{figure}
\centering
\includegraphics[width=90mm]{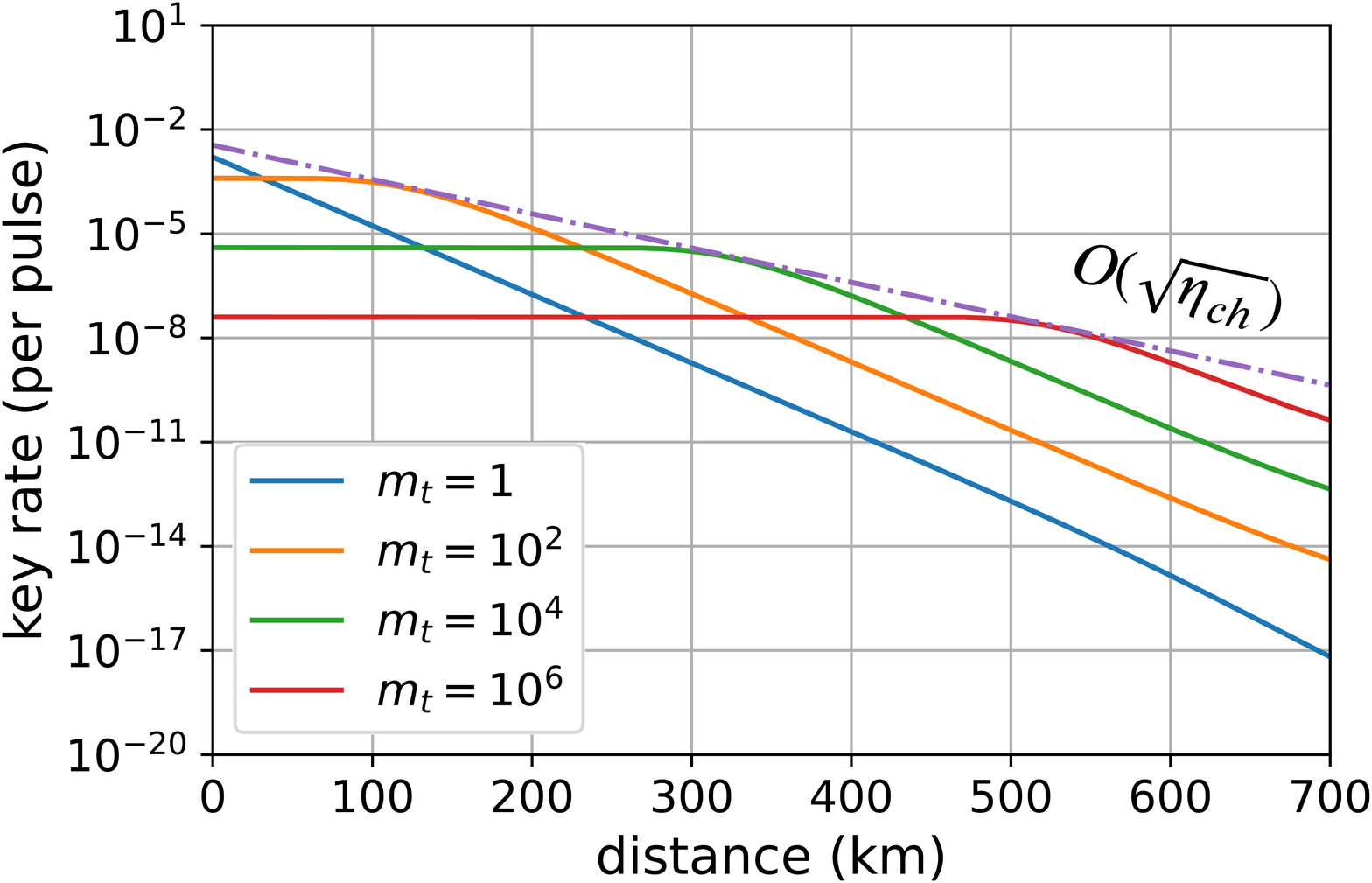}
\caption{Protocol with temporal multiplexing. Simulations were performed with $m_s=1$ in frequency multiplexing and $m_t=1,10^2,10^4,10^6$ for the number of modes in temporal multiplexing.}
\label{fig:afc-ms-mt-protocol}
\end{figure}

As illustrated in Fig.\ref{fig:afc-ms-mt-protocol}, as $m_t$ increases, the plateau region expands; however, the overall rate decreases linearly with an increase in the time required to execute BSM, which is $m_t\tau$. Additionally, optimizing the number of temporal modes $m_t$ with consideration of the distance results in rate scaling of $O(\sqrt{\eta_{ch}})$, as observed in the envelope.

\end{appendices}


\bibliography{sn-bibliography}

\end{document}